\documentclass[11pt]{revtex4}
\usepackage{color}
\usepackage{amsmath,amssymb}
\usepackage[dvipdfm]{graphicx}
\usepackage{cancel}

\allowdisplaybreaks

\begin{document}
\title{Bosonization, Painlev\'{e} property, exact solutions for ${\cal N} =1$ supersymmetric mKdV equation}

\author{Bo Ren$^{1}$\footnote{Electronic mail: renbosemail@gmail.com.}, Jian-Rong Yang$^{2}$, Ping Liu$^{3}$, Xi-Zhong Liu$^{1}$}

\affiliation{$^1$Institute of Nonlinear Science, Shaoxing University, Shaoxing 312000, China \\
$^2$Department of Physics and Electronics, Shangrao Normal University, Shangrao 334001, China \\
$^3$College of Electron and Information Engineering, University of Electronic Science and Technology of China Zhongshan Institute, Zhongshan 528402, China }

\date{\today $\vphantom{\bigg|_{\bigg|}^|}$}

\begin{abstract}
The ${\cal N} =1$ supersymmetric modified Korteweg-de Vries (SmKdV) system is transformed to a system of coupled bosonic equations with the bosonization approach. The bosonized SmKdV (BSmKdV) passes the Painlev\'{e} test and allows a set of B\"{a}cklund transformation (BT) by truncating the series expansions of the solutions about the singularity manifold.
The traveling wave solutions of the BSmKdV system are obtained using the mapping and deformation method. Some special types of exact solutions for the BSmKdV system are found with the solutions and symmetries of the usual mKdV equation. In the meanwhile, the similarity reduction solutions of the system are investigated by using the Lie point symmetry theory. The generalized $\tanh$ function expansion method for the
BSmKdV system leads to a nonauto-BT theorem. Using the nonauto-BT theorem, the novel exact explicit solutions of the BSmKdV system can be obtained. All these solutions obtained via the bosonization procedure are different from those obtained via other methods.

\end{abstract}

\maketitle

\section{Introduction}
The mathematical formulation of supersymmetry is based on the introduction
of Grassmann variables. It exists an extensive literature devoted to
construction the symmsymmetric integrable models, such as Korteweg-de Vries \cite{Kupershmidt}, modified Korteweg-de Vries \cite{Gur,Gurs,Mathieu,Yamanaka}, Sine-Gorden \cite{Hruby}, Kadomtsev-Petviashvili hierarchy \cite{Martin} and nonlinear Sch\"{o}rdinger equation \cite{Roelofs}. It has shown that these supersymmetric integrable systems possess the Painlev\'{e} property, the Lax representation, an infinite number of conservation laws, the B\"{a}cklund and the Darboux transformations, bilinear forms and multi-soliton solutions \cite{Chaichan,Itaru,qpl,qpliu,qplhu,Kersten,Bellucci,Carstea}.
However, to treat the integrable systems with fermions such as the supersymmetric integrable systems and pure integrable fermionic systems is
much more complicated than to study the integrable pure bosonic systems \cite{Plyushchay}. It is
significant if one can establish a proper bosonization procedure to deal with the supersymmetric
systems. Recently, a simple bosonization approach
to treat the super integrable systems has been proposed \cite{Andrea,lou}. The method can effectively avoid difficulties caused by intractable
fermionic fields which are anticommuting \cite{Andrea,lou,renb,Boson}.

In this letter, we shall use the bosonization approach to the SmKdV system \cite{Gur,Gurs,Mathieu,Yamanaka}. It reads
\begin{equation}\label{smkdv}
\Phi_{t} + D^6\Phi - 3 \Phi D^3 \Phi D \Phi - 3 (D \Phi)^2 D^2 \Phi = 0,
\end{equation}
where $D = \partial_\theta + \theta\partial_x$ is the covariant derivative.
It is established with the usual independent variable $\{x, t\}$ and a Grassmann variable $\theta$, and expansion $\Phi$ in terms of $\theta$ yields $\Phi(\theta, x, t)= \xi(x, t) + \theta u(x, t).$ \eqref{smkdv} is related to the ${\cal N} = 1$ supersymmetric KdV through a Miura type of transformation \cite{Mathieu,Yamanaka} and has a bilinear B\"{a}klund transformation \cite{qplhu}. It
shares the common conserved quantities with the supersymetric Sine-Gordon equation \cite{Itaru}. The quasi-periodic wave solutions are constructed with the Hirota bilinear method and the Riemann theta function recently \cite{Quasi}.

The paper is organized as follows. In section 2, based on the bosonization approach, the ${\cal N}=1$ SmKdV system
is changed to a system of coupled bosonic equations. The Painlev\'{e} property
and the BT of the coupled bosonic equations are studied by the standard singularity analysis.
In sections 3, some special types of exact solutions can be
explicitly found by means of the mapping and deformation method.
In sections 4, the reduction solutions for the usual Painlev\'{e} II are found using the Lie point symmetry.
Section 5 is devoted to the generalized $\tanh$ function expansion approach for the coupled bosonic equations.
The explicit novel exact solution of the BSmKdV is investigated. The last section is a simple summary and discussion.

\section{Bosionization of the SmKdV equation and Painlev\'{e} analysis}

\subsection{Bosonization approach with two fermionic parameters}

In terms of the component fields, \eqref{smkdv} is equivalent to
\begin{subequations}\label{com}
\begin{eqnarray}
& u_{t} + u_{xxx} - 6 u^2 u_{x} - 3 \xi (u \xi_x)_x =  0, \\
& \xi_{t}+ \xi_{xxx} - 3u u_x \xi -3u^2\xi_x= 0.
\end{eqnarray}
\end{subequations}
It is obvious that \eqref{com} includes a commuting $u$ and an anticommuting $\xi$ field. It will degenerates to the usual classical system with vanishing the fermionic sector. In order to avoid the difficulties in dealing with the anticommutative fermionic field $\xi$, we expand the component fields $\xi$ and $u$ by introducing the two fermionic parameters \cite{Andrea,lou,renb,Boson}
\begin{subequations}\label{ffield}
\begin{eqnarray}
& u(x, t) = v + w \zeta_1 \zeta_2,\\
& \xi(x, t)= p \zeta_1 + q \zeta_2,
\end{eqnarray}
\end{subequations}
where $\zeta_1$ and $\zeta_2$ are two Grassmann parameters, while the coefficients $v$, $p$, $q$ and $w$ are four usual real or complex functions with respect to the spacetime variable $\{x, t\}$. Substituting \eqref{ffield} into the SmKdV system \eqref{com}, we obtain
\begin{subequations}\label{ffom}
\begin{eqnarray}
& v_{t} + v_{xxx} - 6 v^2 v_{x} =  0, \\
& p_{t} + p_{xxx} - 3 v^2 p_{x} - 3 p v v_{x} = 0, \\
& q_{t} + q_{xxx} - 3 v^2 q_{x} - 3 q v v_{x} = 0, \\
& w_{t} + w_{xxx} - 6 (v^2 w)_x - 3 v_{x} (p q)_x - 3vpq_{xx} - 3vqp_{xx} =  0.
\end{eqnarray}
\end{subequations}
The above way is just the bosonic procedure for the SmKdV system \eqref{com} with two fermionic parameters (BSmKdV-2). (4a) is exactly the usual mKdV equation which has been widely studied \cite{Tanaka,RHirota,Wadati,Yanz}. (4b)-(4d) are linear homogeneous in $p$, $q$ and $w$, respectively. These pure bosonic systems can be easily solved theoretically. This is just one of the advantages of the bosonization approach.

\subsection{Painlev\'{e} analysis and B\"{a}cklund transformations for the BSmKdV-2 system}

In this part, we will study the Painlev\'{e} property and the BT of the BSmKdV-2 system.
If all the movable singularities of its solutions are only poles, the model is called Painlev\'{e} integrable. In order to perform the Painlev\'{e} analysis, the bosonic fields $v, p, q, w$ expand about the singularity manifold $\phi(x, t)=0$ as
\begin{align}\label{trans}
v =\sum^{\infty}_{j=0}v_{j} \phi^{j-\alpha_1}, \hspace{0.6cm} p =\sum^{\infty}_{j=0}p_{j} \phi^{j-\alpha_2}, \hspace{0.6cm} q =\sum^{\infty}_{j=0}q_{j} \phi^{j-\alpha_3}, \hspace{0.6cm} w =\sum^{\infty}_{j=0}w_{j} \phi^{j-\alpha_4},
\end{align}
with $\{v_{j}, p_{j}, q_{j}, w_{j}\}$ being arbitrary functions of $\{x,t\}$.
From the leading order analysis result, the all constants $\alpha_1$, $\alpha_2$, $\alpha_3$ and $\alpha_4$ are positive integers,
i.e., $1, 1, 1, 2$ respectively. Consequently, the recursion relations to determine the functions $v_j$, $p_j$, $q_j$ and $w_j$ can be obtained, the resonance values of $j$ are given
\begin{align}
j= -1, 0, 0, 0, 2, 2, 3, 4, 4, 4, 4, 5.
\end{align}
After the detailed calculations, the resonance conditions
are satisfied identically because the functions $v_{j}$, $p_{j}$, $q_{j}$ and $w_{j}$ are all determined by twelve arbitrary functions $\phi$, $p_0$, $q_0$, $w_0$, $p_2$, $q_2$, $v_3$, $v_4$, $p_4$, $q_4$, $w_4$ and $w_5$. From the above considerations we deduce that the BSmKdV-2 is
really Painlev\'{e} integrable.

Using the standard truncated Painlev\'{e} expansion, the BT is
\begin{align}\label{rom}
v = \frac{v_0}{\phi} + v_1, \hspace{0.6cm} p = \frac{p_0}{\phi} + p_1, \hspace{0.6cm} q = \frac{q_0}{\phi} + q_1, \hspace{0.6cm} w = \frac{w_0}{\phi^2} + \frac{w_1}{\phi} + w_2,
\end{align}
where $v_{0} = \pm \phi_x, \,\,\, w_{1} = (\phi_x w_{0,x} - \phi_{xx}w_0)\phi_x^{-2}$
and $\{v_1, p_1, q_1, w_2\}$ satisfy BSmKdV-2 system.
Besides, we find the fields $\phi$, $p_0$, $q_0$ and $w_0$ are the solutions of the following Schwarzian BSmKdV-2 system
\begin{subequations}\label{schf}
\begin{eqnarray}
&\phi_t + \phi_{xxx} - \frac{3}{2} \frac{\phi_{xx}^2}{\phi_x}=0,  \\
& p_{0,t} + p_{0,xxx} + \frac{15}{4} p_{0,x} \frac{\phi_{xx}^2}{\phi_x^2} + \frac{3}{2} p_{0} \frac{\phi_{xx}\phi_{xxx}}{\phi_x^2} - \frac{3}{2} p_{0,x} \frac{\phi_{xxx}}{\phi_x}  - \frac{9}{4} p_{0} \frac{\phi_{xx}^3}{\phi_x^3} - 3 p_{0,xx} \frac{\phi_{xx}}{\phi_x} = 0,  \\
& q_{0,t} + q_{0,xxx} + \frac{15}{4} q_{0,x} \frac{\phi_{xx}^2}{\phi_x^2} + \frac{3}{2} q_{0} \frac{\phi_{xx}\phi_{xxx}}{\phi_x^2} - \frac{3}{2} q_{0,x} \frac{\phi_{xxx}}{\phi_x}  - \frac{9}{4} q_{0} \frac{\phi_{xx}^3}{\phi_x^3} - 3 q_{0,xx} \frac{\phi_{xx}}{\phi_x} = 0, \\
& w_{0,t} + w_{0,xxx} + \frac{27}{2} w_{0,x} \frac{\phi_{xx}^2}{\phi_x^2} + 6 w_{0} \frac{\phi_{xx}\phi_{xxx}}{\phi_x^2} -
12 w_0 \frac{\phi_{xx}^3}{\phi_x^3} - 3 w_{0,x} \frac{\phi_{xxx}}{\phi_x} - 6 w_{0,xx} \frac{\phi_{xx}}{\phi_x}
+ 3 \frac{q_0 p_{0,xx}\phi_{xx}}{\phi_x} \\
& - 3 \frac{p_0 q_{0,xx}\phi_{xx}}{\phi_x} + \frac{3}{2} \frac{p_0 q_{0,x}\phi_{xxx}}{\phi_x}
- \frac{3}{2} \frac{q_0 p_{0,x}\phi_{xxx}}{\phi_x} + \frac{9}{4} \frac{p_0 q_{0,x}\phi_{xx}^2}{\phi_x^2}
- \frac{9}{4} \frac{q_0 p_{0,x}\phi_{xx}^2}{\phi_x^2} + 3 p_{0,x}q_{0,xx} - 3 q_{0,x}p_{0,xx} =0, \nonumber
\end{eqnarray}
\end{subequations}
with the solutions $\phi$, $p_0$, $q_0$ and $w_0$ are related by
\begin{align}\label{som}
&v_{1}= - \frac{\phi_{xx}}{2\phi_x}, \hspace{1.2cm} p_{1} =
\frac{p_{0}\phi_{xx} - 2 p_{0,x}\phi_{x}}{2 \phi_x^2},\hspace{1.2cm} q_{1} =
\frac{q_{0}\phi_{xx} - 2 q_{0,x}\phi_{x}}{2 \phi_x^2}, \nonumber \\
&w_{2} = \frac{1}{12\phi_x^4}
(12 q_{0}\phi_{xx} p_{0,x} \phi_{x} - 12 p_{0}\phi_{xx} q_{0,x} \phi_{x} + 6 p_{0}q_{0,xx} \phi_{x}^2 - 6 q_{0}p_{0,xx} \phi_{x}^2 \nonumber \\ &
-2w_{0}\phi_t\phi_x - 12 p_{0}\phi_{x}^4 q_{2} + 12 q_{0}\phi_{x}^4 w_0 + 6 w_{0,xx} \phi_{x}^2 - 18 w_{0,x} \phi_{x}\phi_{xx} + 21w_{0}\phi_{xx}^2). \nonumber
\end{align}
It is obvious that an auto-BT \eqref{rom} and a nonauto-BT \eqref{schf} are obtained with the singularity analysis.

\section{Traveling wave solutions with mapping and deformation method}

Now the traveling wave solutions of the bosonic \eqref{ffom} will be studied.
Introducing the traveling wave variable $X=k x+\omega t + c_0$ with constants $k$, $\omega$
and $c_0$, \eqref{ffom} is transformed to the ordinary differential equations (ODEs)
\begin{subequations}\label{fourcom}
\begin{align}
& k^3 v_{XXX} + \omega v_{X} - 6 k v^2 v_{X}= 0, \\
& k^3 p_{XXX} + \omega p_{X} - 3 k pvv_{X} - 3 k v^2 p_{X} = 0,\\
& k^3 q_{XXX} + \omega q_{X} - 3 k qvv_{X} - 3 k v^2 q_{X} = 0,\\
& k^3 w_{XXX} + \omega w_{X} - 6 k (v^2 w)_X + 3 k^2q (v p_{X})_X - 3 k^2p (v q_{X})_X= 0.
\end{align}
\end{subequations}
As the well known exact solutions of (9a), we try to build the mapping and deformation relationship between the traveling wave solutions $v$ and
$\{p, q, w\}$, then the exact solutions of the BSmKdV-2 equation can be obtained with the known solutions of mKdV equation.

At first, we get $v_{X}$ from (9a)
\begin{align}\label{u0}
v_{X} = \frac{a_0 \sqrt{k(kv^4-\omega v^2 - 2c_1v + c_2k^3)}}{k^2},
\end{align}
where $c_1$ and $c_2$ are the integral constants and $a_0^2=1$.
In order to get the mapping relationship between $v$ and $\{p, q, w\}$, we introduce the variable transformations
\begin{align}\label{trans}
p(X)=P(v(X)), \hspace{0.8cm} q(X)=Q(v(X)), \hspace{0.8cm} w(X)=W(v(X)).
\end{align}
Using the transformation \eqref{trans} and vanishing $v_{X}$ via \eqref{u0}, the linear ODEs (9b)-(9d) become
\begin{subequations}\label{map}
\begin{align}
& \bigl(k v^4 - \omega v^2 - 2c_1v + c_2k^3 \bigr) \frac{\emph{d}^3 P}{\emph{d} v^3} + \bigl(6k v^3 - 3\omega v - 3c_1 \bigr) \frac{\emph{d}^2 P }{\emph{d} v^2} + 3k v^2 \frac{\emph{d} P}{\emph{d} v} - 3 k v P =0, \\
& \bigl(k v^4 - \omega v^2 - 2c_1 v + c_2k^3 \bigr) \frac{\emph{d}^3 Q}{\emph{d} v^3} + \bigl(6k v^3 - 3\omega v - 3c_1 \bigr) \frac{\emph{d}^2 Q }{\emph{d} v^2} + 3kv^2 \frac{\emph{d} Q }{\emph{d} v} - 3 k v Q=0, \\
& \bigl(k v^4 - \omega v^2 - 2c_1v + c_2k^3 \bigr) \frac{\emph{d}^2 W}{\emph{d} v^2} + \bigl(2k v^3 - \omega v - c_1 \bigr) \frac{\emph{d} W }{\emph{d} v} + (\omega - 6kv^2) W - F(v)=0,
\end{align}
\end{subequations}
where $$F(v)= 3a_0\sqrt{k(k v^4 - \omega v^2 - 2c_1 v + c_2k^3 )}\Bigl(P\frac{\emph{d} Q}{\emph{d} v} - Q\frac{\emph{d} P}{\emph{d} v}\Bigr) +c_3.$$
The mapping and deformation relations are constructed via \eqref{map}
\begin{subequations}\label{mapp}
\begin{align}
& P = A_1 v + (A_2+A_3)v \cos(R(v)) - (A_2-A_3)v \sin(R(v)),\\
& Q = A_4 v + (A_5+A_6)v \cos(R(v)) - (A_5-A_6)v \sin(R(v)),\\
& W = \biggl(A_7 + \int^{v} \frac{A_8 + y F(y)}{(k y^4 - \omega y^2 - 2c_1 y + c_2k^3)^{3/2}}dy\biggr)\sqrt{k v^4 - \omega v^2 - 2c_1 v + c_2k^3},
\end{align}
\end{subequations}
where $A_k \,(k=1, 2, \cdots, 8)$ are arbitrary constants,
and $$R(v)=\int^u\frac{ik\sqrt{c_2k}}{{y\sqrt{k y^4 - \omega y^2 - 2c_1 y + c_2k^3}}}dy.$$
If we know the solution of $v$, the traveling wave solution of BSmKdV-2 system will be given with considering \eqref{mapp} and \eqref{trans}.
For a special case, $A_8 =c_3= 0$, $A_2 =A_5$ and $A_3 =A_6$, the traveling wave solution $W$ is
an ordinary type of the symmetries of the traveling wave equation (9a).
In fact, for any given a solution $v$ of the usual mKdV equation, a special type solutions of the bosonic equation \eqref{ffom} can be constructed
\begin{align}
p=P=A_1v, \hspace{1cm} q=Q=A_4v, \hspace{1cm} w=W= A_{7} \sigma(v),
\end{align}
where $\sigma(v)$ is the symmetry of the usual mKdV equation (9a).
The field $w$ exactly satisfy the symmetry equation of the usual mKdV system. The solution $v$ is not restricted to the traveling
wave solutions. We can construct not only traveling
wave solutions but also some novel types of solutions of the
BSmKdV-2 system by using the solutions and symmetries of the mKdV equation.

\section{Similarity reduction solutions with Lie point symmetry theory}

It is well known that the Lie point symmetry play an important role in the investigation
of nonlinear partial differential equations (PDEs) in modern mathematical physics.
The approach is effective methods to obtain the
explicit exact solutions~\cite{Olver,ss8,ss9,rens}. Our aim is to apply the techniques of Lie group theory to the coupled bosonic equation in order
to obtain particular exact solutions and to study their properties.
First, we assume the corresponding Lie point symmetry has the vector form
\begin{align}
V = X \frac{\partial}{\partial x} +  T \frac{\partial}{\partial t} + V \frac{\partial}{\partial v} + P \frac{\partial}{\partial p} +
Q \frac{\partial}{\partial q} + W \frac{\partial}{\partial w},
\end{align}
where $X$, $T$, $V$, $P$, $Q$ and $W$ are functions with respect to $x$, $t$, $v$, $p$, $q$ and $w$. The symmetry supposes
\begin{align}\label{symm}
\sigma_0 = Xv_{x} + Tv_{t} - V, \hspace{0.2cm} \sigma_{1} = Xp_{x} + Tp_{t} - P, \hspace{0.2cm} \sigma_{2} = Xq_{x} + Tq_{t} - Q, \hspace{0.2cm} \sigma_{3} = Xw_{x} + Tw_{t} - W.
\end{align}
The symmetry $\sigma_k \,(k=0,1,2,3)$ is the solution of the linearized equations of \eqref{ffom}
\begin{subequations}\label{fosym}
\begin{eqnarray}
& \sigma_{0,t} + \sigma_{0,xxx} - 6 (v^2 \sigma_{0})_x = 0, \\
& \sigma_{1,t} + \sigma_{1,xxx} - 3 vv_x\sigma_1 - 3 v^2\sigma_{1,x} - 3p (\sigma_{0}v)_x - 6v\sigma_{0}p_x = 0,\\
& \sigma_{2,t} + \sigma_{2,xxx} - 3 vv_x\sigma_2 - 3 v^2\sigma_{2,x} - 3q (\sigma_{0}v)_x - 6v\sigma_{0}q_x = 0,\\
& \sigma_{3,t} + \sigma_{3,xxx} - 6 (\sigma_{3} v^2)_x - 12 (\sigma_0 v w)_x + 3 \sigma_{2}(vp_x)_x - 3 \sigma_{1} (vq_x)_x
\nonumber \\ & - 3 p (\sigma_{0}q_x+\sigma_{2,x}v)_x + 3 q (\sigma_{0}p_x+\sigma_{1,x}v)_x =0.
\end{eqnarray}
\end{subequations}
Substituting \eqref{symm} into the symmetry equations \eqref{fosym} and eliminating $v$, $p$, $q$ and $w$ in terms of \eqref{ffom},
the solutions $X$, $T$, $V$, $P$, $Q$ and $W$ can be concluded using the determining equations
\begin{align}\label{symso}
&T = C_1 t + C_2, \hspace{1.2cm} X= \frac{C_1x}{3} + C_7, \hspace{1.2cm} V=-\frac{C_1}{3}v, \\  \nonumber
&P=C_3 p + C_4 q,  \hspace{1.1cm} Q= C_{6} q + C_{5} p,  \hspace{1.1cm} W = w(C_{3} + C_{6}),
\end{align}
where $C_i \,(i=1,2,...,7)$ are arbitrary constants.
Then, one can solve the characteristic equations
\begin{align}
\frac{\emph{d} x}{X}=\frac{\emph{d}t}{T}, \hspace{1cm} \frac{\emph{d} v}{V}=\frac{\emph{d}t}{T}, \hspace{1cm}
\frac{\emph{d} p}{P}=\frac{\emph{d}t}{T}, \hspace{1cm} \frac{\emph{d} q}{Q}=\frac{\emph{d}t}{T}, \hspace{1cm} \frac{\emph{d} w}{W}=\frac{\emph{d}t}{T},
\end{align}
where $X$, $T$, $V$, $P$, $Q$, and $W$ are given by \eqref{symso}.
One case about the solution \eqref{ffom} will discuss in the following.

When $C_1=C_4=C_5=0$, we can find the similarity solutions after solving out the characteristic equations
\begin{align}\label{slotu}
v=V(\xi), \hspace{0.8cm} p=P(\xi)e^{\frac{C_3x}{C_7}},\hspace{0.8cm} q=Q(\xi)e^{\frac{C_6x}{C_7}},\hspace{0.8cm} w=W(\xi)e^{\frac{(C_3+C_6)x}{C_7}},
\end{align}
with the similarity variable $\xi=t-\bigl(C_2/C_7\bigr)x$. Substituting \eqref{slotu} into \eqref{ffom}, the invariant functions $V$, $P$, $Q$ and $W$ satisfy the reduction systems
\begin{subequations}\label{similarity}
\begin{align}
& V_{\xi\xi\xi} - \frac{C_7^3}{C_2^3} V_{\xi} - \frac{6C_7^2}{C_2^2} V^2 V_{\xi} =0,\\
& P_{\xi\xi\xi} - \frac{3C_3}{C_2} P_{\xi\xi} - \frac{3C_7^2}{C_2^2}P_\xi V^2 - \frac{3C_7^2}{C_2^2} PVV_\xi + \frac{3C_2C_3^2-C_7^3}{C_2^3} P_\xi V^2 - \frac{C_3^3}{C_2^3} P + \frac{3C_3C_7^2}{C_2^3} P V^2 =0,\\
& Q_{\xi\xi\xi} - \frac{3C_6}{C_2} Q_{\xi\xi} - \frac{3C_7^2}{C_2^2}Q_\xi V^2 - \frac{3C_7^2}{C_2^2} QVV_\xi + \frac{3C_2C_6^2-C_7^3}{C_2^3} Q_\xi V^2 - \frac{C_6^3}{C_2^3} Q + \frac{3C_6C_7^2}{C_2^3} Q V^2 =0\\
& W_{\xi\xi\xi} + \frac{C_3^3-3C_2^2C_6-3C_2^2C_3}{C_2^3} W_{\xi\xi} - \frac{6C_7^2}{C_2^2} (V^2 W)_{\xi} - \frac{C_3^3+C_6^3+3C_3C_6^2+3C_3^2C_6}{C_2^3} W \nonumber \\
& + \frac{6C_7^2(C_3+C_6)}{C_2^3} V^2 W + \frac{3C_2C_6^2+3C_2C_3^2+6C_2C_3C_6-C_7^3}{C_2^3} W_\xi + \frac{3C_7(C_6^2-C_3^2)}{C_2^3} VPQ \\
& - \frac{3C_7}{C_2} Q(VP_\xi)_\xi + \frac{3C_7}{C_2} P(VQ_\xi)_\xi + \frac{3C_7(C_3-C_6)}{C_2^2}V_\xi PQ + \frac{6C_7}{C_2^2} V (C_3P_\xi Q - C_6PQ_\xi) =0. \nonumber
\end{align}
\end{subequations}
which (21a) satisfies the Painlev\'{e} II equation. These reduction equations are linear ODEs while the previous functions are known, we can solve $V$, $P$, $Q$, and $W$ one after another in principle. The group invariant solution
of an interaction solution among a Painlev\'{e} II wave and a soliton is obtained with the Lie point symmetry theory.

\section{Generalized tanh function expansion method of BSmKdV-2 system}

The truncated Painlev\'{e} expansion approach and the generalized tanh function expansion method are established to find interactions
among different nonlinear excitations \cite{Interactions}. The methods are valid
for all integrable systems and or even nonintegrable models because both the truncated Painlev\'{e} analysis
and the tanh expansion method can be used to find exact solutions of the partially solvable nonlinear models \cite{Interactions,Boson}.

According to the usual tanh function expansion method, the generalized expansion solution has the form
\begin{subequations}\label{ff}
\begin{align}
& v = v_{0} + v_{1}\tanh (f), \\
& p = p_{0} + p_{1}\tanh (f),\\
& q = q_{0} + q_{1}\tanh (f),\\
& w = w_{0} + w_{1}\tanh (f) + w_{2}\tanh^2(f).
\end{align}
\end{subequations}
where $v_{0}$, $v_{1}$, $p_{0}$, $p_{1}$, $q_{0}$, $q_{1}$, $w_{0}$, $w_{1}$, $w_2$ and $f$ are functions of
$\{x, t\}$ and should be determined later.
After some detail calculations by substituting \eqref{ff} into the BSmKdV-2 system \eqref{ffom},
we can prove the following nonauto-BT theorem.
\\
{\bf Theorem (Nonauto-BT theorem).} If $\{f,\, g,\, h,\, n\}$ is a solution of
\begin{subequations}\label{fvet}
\begin{align}
& f_t + f_{xxx} - 2 f_x^3 - \frac{3}{2} \frac{f_{xx}^2}{f_x} =0, \\
& g_t + g_{xxx} - 6 g_x f_x^2 - g_{xx}\frac{3 f_{xx}}{f_x} + g_x \frac{3f_t}{2f_x} + g_x \frac{3f_{xx}^2}{2f_x^2}- g \frac{3f_tf_{xx}^2}{2f_x^2}=0,\\
& h_t + h_{xxx} - 6 h_x f_x^2 - h_{xx}\frac{3 f_{xx}}{f_x} + h_x \frac{3f_t}{2f_x} + h_x \frac{3f_{xx}^2}{2f_x^2}- h \frac{3f_tf_{xx}^2}{2f_x^2}=0,\\ \nonumber
& n_t + n_{xxx} - n_{xx} \frac{6f_{xx}}{f_x} + n_x \bigl(\frac{9f_{xx}^2}{f_{x}^2} +\frac{3f_t}{f_x} -12 f_x^2\bigr) - n \bigl(\frac{6f_tf_{xx}}{f_x^2} + \frac{3f_{xx}^3}{f_x^3}-12 f_xf_{xx}\bigr) + 6f_x^2 (gh_x - hg_x) \nonumber  \\
& + \frac{3f_t}{2f_x} (hg_x - gh_x)
 + \frac{9f_{xx}^2}{2f_x^2} (hg_x - g h_x) + \frac{3f_{xx}^2}{f_x} (h g_{xx} - 3g h_{xx})+ 3g_xh_{xx}- 3h_xg_{xx}=0.
\end{align}
\end{subequations}
then $\{v,\, p,\, q,\, w\}$ with
\begin{subequations}\label{fvetor}
\begin{align}
v & = -\frac{f_{xx}}{2f_x} + f_x \tanh(f), \\
p & = \frac{f_{xx}}{2f_x^2}g -\frac{g_x}{f_x} + g\tanh(f),\\
q & = \frac{f_{xx}}{2f_x^2}h -\frac{h_x}{f_x} + h\tanh(f),\\ \nonumber
w & =  \frac{n_{xx}}{2f_x^2} - n_x\frac{3f_{xx}}{2f_x^3} + n(\frac{f_t}{2f_x^3}+\frac{3f_{xx}^2}{4f_x^4}-2) + \frac{1}{2f_x^2}(gh_{xx} - hg_{xx}) + \frac{f_{xx}}{f_x^3} (hg_x - gh_x) \\
& + (\frac{f_{xx}n}{f_x^2} - \frac{n_x}{f_x})\tanh(f) + n\tanh^2(f).
\end{align}
\end{subequations}
is a solution of the BSmKdV-2 system \eqref{ffom}.

We can find some nontrivial solutions of the BSmKdV-2 from some quite
trivial solutions of \eqref{fvet}. Here we list an interesting examples.
A quite trivial straight-line solution of \eqref{fvet} has the form
\begin{align}\label{solm}
& f = k_0 x+\omega_0 t + l_0, \hspace{0.6cm} g = k_1 x+\omega_1 t + l_1, \hspace{0.6cm} h = k_2 x+\omega_2 t + l_2, \hspace{0.6cm} n = k_3 x+\omega_3 t + l_3, \nonumber\\
& \omega_0=2k_0^3, \hspace{1.2cm} \omega_1 = 3k_0^2k_1, \hspace{1.2cm} \omega_2 = 3k_0^2k_2, \hspace{1.2cm}
\omega_3 = 3 k_0^2 (2 k_3 + k_1 l_2 - k_2 l_1),
\end{align}
where $k_0$, $k_1$, $k_2$, $k_3$, $l_0$, $l_1$, $l_2$ and $l_3$ are all the free constants. Substituting the line solution \eqref{solm} into the nonauto-BT theorem yields the following soliton solution of BSmKdV-2 system
\begin{subequations}\label{fsolu}
\begin{align}
v & = k_0 \tanh(f), \\
p & = g\tanh(f)-\frac{k_1}{k_0},\\
q & = h\tanh(f)-\frac{k_2}{k_0},\\
w & = n\tanh^2(f)-\frac{k_3}{k_0}\tanh(f) - n.
\end{align}
\end{subequations}
Though the soliton solution \eqref{fsolu} is a traveling wave in the space time $\{x, t\}$ for the boson
field $v$, it is not a traveling wave for other boson fields $p$, $q$ and $w$, then the superfiled
$\Phi$ of SmKdV is not a traveling wave except for the case of $g$, $h$ and $n$ being constants, i.e.,
$k_1=k_2=k_3=0$. This example reveals that an straightening the single soliton to a straight-line solution for the
BSmKdV-2 is given by the nonauto-BT theorem.

\section{Conclusions}

In summary, the bosonization procedure has been successfully applied to the SmKdV equation. The SmKdV equation is simplified to the mKdV equation together with three linear differential equations. The BSmKdV-2 system is proved to possess Painlev\'{e} property and to be completely integrable. The auto-BT and nonauto-BT are constructed by truncating the standard Painlev\'{e} expansion.

The traveling wave solutions are studied by using the mapping and deformation method. Some special types of exact solutions can be given straightforwardly through the exact solutions of the mKdV equation and its symmetries.
In addition, the group invariant solutions of the system are derived with the Lie point symmetry method.
The generalized tanh function expansion method is developed to find interaction solutions among different nonlinear excitations.
Straightening a single soliton to a straight-line solution for the BSmKdV-2 system is constructed with the generalized tanh function expansion method. Using the nonauto-BT theorem, various exact explicit solutions of the BSmKdV-2 system can be obtained.
All these solutions obtained via the bosonization procedure are different from those obtained
via other methods such as the Hirota bilinear method and the Riemann theta function \cite{qplhu,Quasi}.

In this paper, the properties and exact solutions of the BSmKdV-2 system are investigated,
we can also introduce $N$ fermionic parameters to expand the SmKdV system (BSmKdV-N).
For the $N\geq 2$ fermionic parameters $\zeta_i \,(i=1,2,...,N)$ instance, the component fields $u$ and $\xi$ expand
\addtocounter{equation}{1}
\begin{equation}
\xi(x, t)= \sum _{n=1}^{[\frac{N+1}{2}]}\sum_{1\leq i_1 <\cdots< i_{2n-1} \leq N} u_{i_1i_2\cdots i_{2n-1}}\zeta_{i_1}\zeta_{i_2}\cdots\zeta_{i_{2n-1}}, \tag{\theequation a}
\end{equation}
\begin{equation}
u(x, t) = u_0 + \sum _{n=1}^{[\frac{N+1}{2}]}\sum_{1\leq i_1 <\cdots< i_{2n} \leq N} u_{i_1i_2\cdots i_{2n}}\zeta_{i_1}\zeta_{i_2}\cdots\zeta_{i_{2n}}, \tag{\theequation b}
\end{equation}
where the coefficients $u_0,  u_{i_1i_2\cdots i_{2n}} \,(1\leq i_1 <\cdots< i_{2n} \leq N)$ and $u_{i_1i_2\cdots i_{2n-1}} \,(1\leq i_1 <\cdots< i_{2n-1} \leq N)$ are $2^N$ real or complex bosonic functions of classical spacetime variable $\{x, t\}$.
The Painlev\'{e} property and exact solutions are worthy study for the BSmKdV-N system.

{\bf Acknowledgment}:

\noindent
We would like to thank S. Y. Lou for useful discussions.
This work was partially supported by the National Natural Science Foundation of China under Grant (Nos. 11305106, 11365017 and 11305031), the Natural Science Foundation of Zhejiang Province of China under Grant (No. LQ13A050001) and the Natural Science Foundation of Guangdong Province (No. S2013010011546).


\end{document}